\documentclass[lettersize,journal]{IEEEtran}
\usepackage{amsmath,amsfonts}
\usepackage{algorithmicx}
\usepackage{algorithm}
\usepackage{algpseudocode}
\usepackage{changes}
\usepackage{multirow}
\usepackage{array}
\usepackage[caption=false,font=normalsize,labelfont=sf,textfont=sf]{subfig}
\usepackage{textcomp}
\usepackage{stfloats}
\usepackage{url}
\usepackage{verbatim}
\usepackage{graphicx}
\usepackage{cite}
\usepackage{xcolor}
\usepackage{color}
\usepackage{array, makecell}
\usepackage{amssymb}
\usepackage{pifont}
\begin{document}

\title{ApproxABFT: Approximate Algorithm-Based Fault Tolerance for Neural Network Processing}

\author {Xinghua Xue, ~\IEEEmembership{Member,~IEEE},
         Cheng Liu, ~\IEEEmembership{Senior Member,~IEEE},\\
         Feng Min,
         Tao Luo, ~\IEEEmembership{Member,~IEEE,}
         Yinhe Han, ~\IEEEmembership{Senior Member,~IEEE}

\thanks{Cheng Liu is the corresponding author. Xinghua Xue is with Hangzhou Institute for Advanced Study, University of Chinese Academy of Sciences. Cheng Liu is with the State Key Lab of Processors, Institute of Computing Technology, Chinese Academy of Sciences. Feng Min is with Institute of Computing Technology, Chinese Academy of Sciences. Tao Luo is with Institute of High Performance Computing, A*STAR, Singapore. Yinhe Han is with Institute of Computing Technology, Chinese Academy of Sciences. This paper is supported in part by National Key Research and Development Program of China under grant No.(2022YFB3902802) and National Natural Science Foundation of China (NSFC) under grant No.(62174162).}
\thanks{Manuscript received X X, XXX; revised X X, XXX; revised X X, XXX.}}

\markboth{IEEE TRANSACTIONS ON VERY LARGE SCALE INTEGRA TION (VLSI) SYSTEMS, VOL. XX, NO. XX, XXX 2025}%
{Xinghua Xue \MakeLowercase{\textit{et al.}}: Bare Demo of IEEEtran.cls for IEEE Journals}

\maketitle

\begin{abstract}

With the increasing deployment of deep neural networks (DNNs) in terrestrial and aerospace safety-critical applications, system reliability has emerged as a co-equal design metric alongside computational efficiency. Algorithm-based fault tolerance (ABFT) mechanisms, characterized by architecture-agnostic and cost-effectiveness, have become a promising solution for reliability enhancement. However, conventional ABFT approaches rely on rigorous verification mechanisms where even minor computational deviations trigger error recovery processes, which not only disregards the intrinsic fault tolerance characteristics of DNN models but also incurs redundant fault tolerance processing overhead. To address these limitations, we propose an Approximate ABFT framework (ApproxABFT) that innovatively introduces adaptive error tolerance thresholds to enable selective fault recovery, activating error correction modules exclusively when computational deviations exceed predefined thresholds. This approach effectively mitigating overreaction to non-critical computational errors. Furthermore, a dynamic block granularity optimization algorithm is implemented to achieve inter-layer error sensitivity balancing. Experimental evaluations demonstrate that the proposed ApproxABFT achieves a 43.39\% average reduction in redundant computing overhead compared to previous accurate ABFT, while simultaneously enhancing the tolerable soft error rate by an order of magnitude.

\end{abstract}


\section{Introduction} 

Deep Neural Networks (DNNs) are currently extensively utilized in tasks such as autonomous spacecraft navigation and real-time analysis of satellite imagery. In the context of space satellite missions, the occurrence of bit-flip errors induced by high-energy particle radiation constitutes a significant factor that compromises the reliability of the system \cite{aguiar2025single,dilillo2022space,tarakcciouglu2023disruptive}. In contrast to the stable terrestrial environment, the radiation in space exhibits a persistent and unpredictable nature. According to on-orbit observations and terrestrial radiation testing records\cite{mar,CARMEN2/MEX,CARMEN2,NASA,juan}, the probability of single bit flip error in the space environment can range from $10^{-7}$ to $10^{-6}$ per bit per day . This implies that a typical application utilizing 40MB of memory may encounter between 30 to 300 bit-flip errors daily. More critically, approximately 5\% of radiation events can induce multi-bit errors \cite{mult,wang2024case}, posing a threat to memory- or computation-intensive tasks such as the neural networks. Even a single bit error within a computational layer can propagate through calculations, potentially causing deviations in hundreds of output values.

Existing hardware-level protection schemes can enhance the reliability but introduce excessive overhead. For instance, the Dual Modular Redundancy (DMR) and the Triple Modular Redundancy (TMR) require an additional 100\% to 200\% of computational resources\cite{rogenmoser2025hybrid,lyons1962use}, rendering such solutions prohibitively costly for resource-constrained edge satellites. The Error-Correcting Code (ECC) techniques\cite{dopson2005softecc}, while capable of correcting single-bit errors, necessitate 5 bits of redundant information per byte, leading to a 62.5\% increase in memory overhead. Software compiler-level redundancy methods\cite{baroffio2024enhanced} do not require hardware modifications but result in a multiplicative increase in runtime. Against this backdrop, model-level fault tolerance methods\cite{ahmadilivani2024cost,jung2024maintaining} have emerged as a focal point of research, though their applicability in high-radiation scenarios remains limited. Current lightweight approaches, such as activation restriction\cite{mousavi2024proact} and selective replication\cite{xue2023exploring}, can reduce overhead but sacrifice the expression capabilities of the model. Moreover, fault-aware training schemes\cite{dutta2019codenet,pourmehrani2024fat}, which essentially learn fault information to adapt to specific fault scenarios, suffer from a lack of generalization ability in the model.

Algorithm-based fault tolerance(ABFT) \cite{huang1984algorithm} employs embedded mathematical verification mechanisms to detect and recover from computational errors without necessitating modifications to the hardware architecture, which is a breakthrough method due to its low overhead characteristics.  ABFT has been successfully implemented in computational models such as General Matrix Multiplication (GEMM) and convolution operations. Existing ABFT fault tolerance schemes \cite{safarpour2021low, roffe2020evaluation, sharif2023efficient, li2022efficient, zamani2019greenmm, liu2018fault} have demonstrated satisfactory performance in low-fault scenarios. However, they encounter two significant challenges when deployed in high-radiation space environments. Firstly, the frequent execution of error recovery operations can lead to a substantial increase in computing overhead for ABFT. Secondly, the stringent verification mechanism of traditional ABFT fails to maintain fault tolerance when multiple computational errors occur simultaneously in both a row and a column.

We have observed that neural networks demonstrate significant resilience to soft errors, maintaining robust output accuracy even in the presence of numerous minor computational deviations  as shown in Section \ref{sec:motivation}. This phenomenon prompts us to re-examine the previous accurate ABFT approach. Specifically, we explore whether the stringent error detection and recovery mechanisms can be integrated with the inherent fault-tolerant characteristics of neural networks to uncover novel optimization potential. Building on this insight, we innovatively introduce the concept of approximation into accurate ABFT, proposing an Approximate ABFT method, termed ApproxABFT. The breakthrough of this method lies in leveraging the intrinsic fault tolerance of neural networks to relax the thresholds for error detection and recovery in accurate ABFT. This allows numerous minor computational errors to be disregarded, thereby avoiding the high-cost error recovery process. Concurrently, the method effectively identifies and prioritizes significant error deviations, enabling precise allocation of fault tolerance resources to enhance error recovery probability in multi-fault scenarios. Ultimately, this approach achieves the dual objectives of improving model accuracy and reducing fault tolerance overhead. ApproxABFT exhibits strong compatibility with diverse neural network architectures.

In practical applications, the sensitivity of different neural network layers to soft errors differs significantly due to varying accuracy requirements across different scenarios. If the entire model adopts a fixed fault tolerance threshold, it will inevitably result in either excessive computational resource wastage due to over-protection, or diminished fault tolerance capability due to insufficient protection. To address this challenge in ApproxABFT, we design a multi-parameter collaborative optimization algorithm that constructs a dynamic optimization space. This framework enables the joint optimization of error detection approximation thresholds, error recovery approximation thresholds, and block size parameters across different layers. The algorithm adaptively aligns with the varying error sensitivities across different layers of neural networks to improve performance. Experimental results demonstrate that compared with the previous accurate ABFT methods, our ApproxABFT scheme not only achieves significant improvements in model accuracy but also substantially reduces the overhead associated with fault tolerance design. The contributions of this work are summarized as follows.

\begin{itemize}
\item We propose an approximate ABFT (ApproxABFT) , designed to relax the stringent constraints on error detection and recovery inherent in previous accurate ABFT. This approach effectively mitigates costly error recovery processes while enhancing the probability of error recovery in multi-fault scenarios.

\item We investigated the fault tolerance of neural networks and analyzed the distribution of computational deviations induced by soft errors. Our observations reveal that the majority of deviations caused by soft errors in matrix sums and row/column sums are minimal, thereby providing substantial design flexibility for the proposed ApproxABFT.

\item We formulated the threshold optimization problem and proposed a multi-parameter collaborative optimization algorithm to address the intricate design trade-offs among approximation strength, fault tolerance overhead, and model accuracy.

\item According to our experiments,  ApproxABFT achieves 43.39\% reduction in computing overhead compared to previous accurate ABFT, while increasing the tolerable error rate by an order of magnitude on average. 

\end{itemize}

\section{Related Work}

In addressing the issue of silicon-based errors during neural network computations, the academic community has developed a multi-level technical approach. Traditional hardware-level redundancy schemes \cite{barbirotta2022design,aviles2024hardening,vafaei2024hpr} achieve error masking through physical replication but incur multiple-fold hardware resource overhead. ECC technology \cite{feinberg2018making} exhibits the capability to correct single-bit errors, yet its implementation necessitates substantial memory overhead. Particularly when extended to multi-bit error tolerance scenarios, the memory overhead associated with redundant check data further escalates. Software compilation-level redundancy method \cite{kaya2024compiler} reduce error probability through instruction stream replication, but they introduce a multiplicative runtime cost, making them impractical for real-time tasks. Against this backdrop, model-level fault tolerance methods, which decouple from hardware architecture through algorithmic innovation, have gradually become a focal point of research. Activation restriction strategies\cite{zhan2021improving} suppress error propagation by truncating anomalous values, but they sacrifice the nonlinear expression capabilities of the model and exhibit inadequate error recovery in high-radiation environments. Selective replication schemes\cite{xue2022winograd} can provide targeted protection for critical computational nodes, yet their protection mechanisms struggle to adapt to dynamic radiation conditions. Fault-aware training schemes\cite{dos2025improving}, which essentially learn fault patterns to adapt to specific fault scenarios, suffer from inherent limitations such as poor scenario generalization and high retraining costs.

In the course of this technological evolution, ABFT grounded in checksum techniques, has emerged as a promising approach for neural network fault tolerance due to its hardware-agnostic nature and low overhead. Originating from research on resilient computation for GEMM, this technique embeds mathematical verification mechanisms to enable error detection and recovery, thereby circumventing the cost pressures associated with hardware modifications. As the computational paradigms of deep learning have evolved, the application scope of ABFT have continuously expanded, ranging from fault-tolerant implementations of Fast Fourier Transform (FFT) to checksum optimizations for sparse tensor operations, such as FFT\cite{wang1994algorithm,liang2017correcting,wu2024turbofft}, sparse operations\cite{scholl2016efficient,peltekis2024error}, decomposition\cite{wu2016towards, leon2024comparative ,chen2018fault, nguyen2020coded}, and sorting\cite{li2019ft,camargo2024algorithm}. A high-performance FFT implementation \cite{wu2024turbofft} was proposed, equipped with a two-sided fault tolerance scheme that detects and corrects silent data corruptions at computing units on-the-fly. In recent years, ABFT research has expanded into the core convolutional operators of deep learning\cite{zhao2020ft}, systematically constructing a multi-level checksum system for convolutional layers and designing a multi-scheme workflow to achieve high detection capabilities. To significantly reduce execution time overhead, parallelized checksum schemes have been developed\cite{kosaian2021arithmetic}, leveraging the high compute-to-memory bandwidth ratio of GPUs to breakthrough time overhead limitations. To enhance error correction capabilities against faults, a three-level fused ABFT mechanism tailored for GPU architectures has been proposed\cite{wu2023anatomy} , achieving multi-granularity error protection while ensuring computational efficiency.

However, the underlying design paradigm of previous accurate ABFT methods \cite{xue} still remains on the traditional exact verification mechanisms, which perform well in low error rate scenarios. When deployed in sustained high-energy radiation environments, accurate ABFT confronts two critical limitations: firstly, frequent error recovery operations cause its computing overhead to grow quadratically; secondly, when multiple computational errors occur in a row or column, the strict verification mechanism of traditional ABFT leads to a failure in fault tolerance. These issues highlight the contradiction that existing solutions have yet to overcome: how to trade-off between computational efficiency and fault tolerance capability in high-radiation environments.

\section{Observation and Motivation} \label{sec:motivation}

This section investigates the computational error distribution characteristics of previous accurate ABFT under soft-error scenarios, elucidates its correlation mechanism with model accuracy, and reveals critical challenges in fault-tolerant protection implementation.

The VGG19 architecture on the CIFAR10 dataset is selected as the baseline model, and the bit error rate (BER) is used to quantify bit-flip effect intensity. Reliability evaluation experiments are conducted based on the operation-wise fault injection platform \cite{xue2023exploring}, which injects random bit-flip errors into the outputs of primitive operations such as addition and multiplication. In the error detection phase, the Matrix Sum Deviation (MSD) metric is introduced, as defined in Eq. (\ref{eq:detect}), which represents the computational deviation between the predicted matrix checksum and the actual output matrix checksum. For the error recovery phase, the Row/Column Sum Deviation (R/CSD) metric is introduced, as defined in Eq. (\ref{eq:sum}), which represents the computational deviation between the predicted row/column checksum vector and the actual output row/column checksum vector.

\begin{figure}[!t]

\centering
\includegraphics[width=0.47\textwidth]{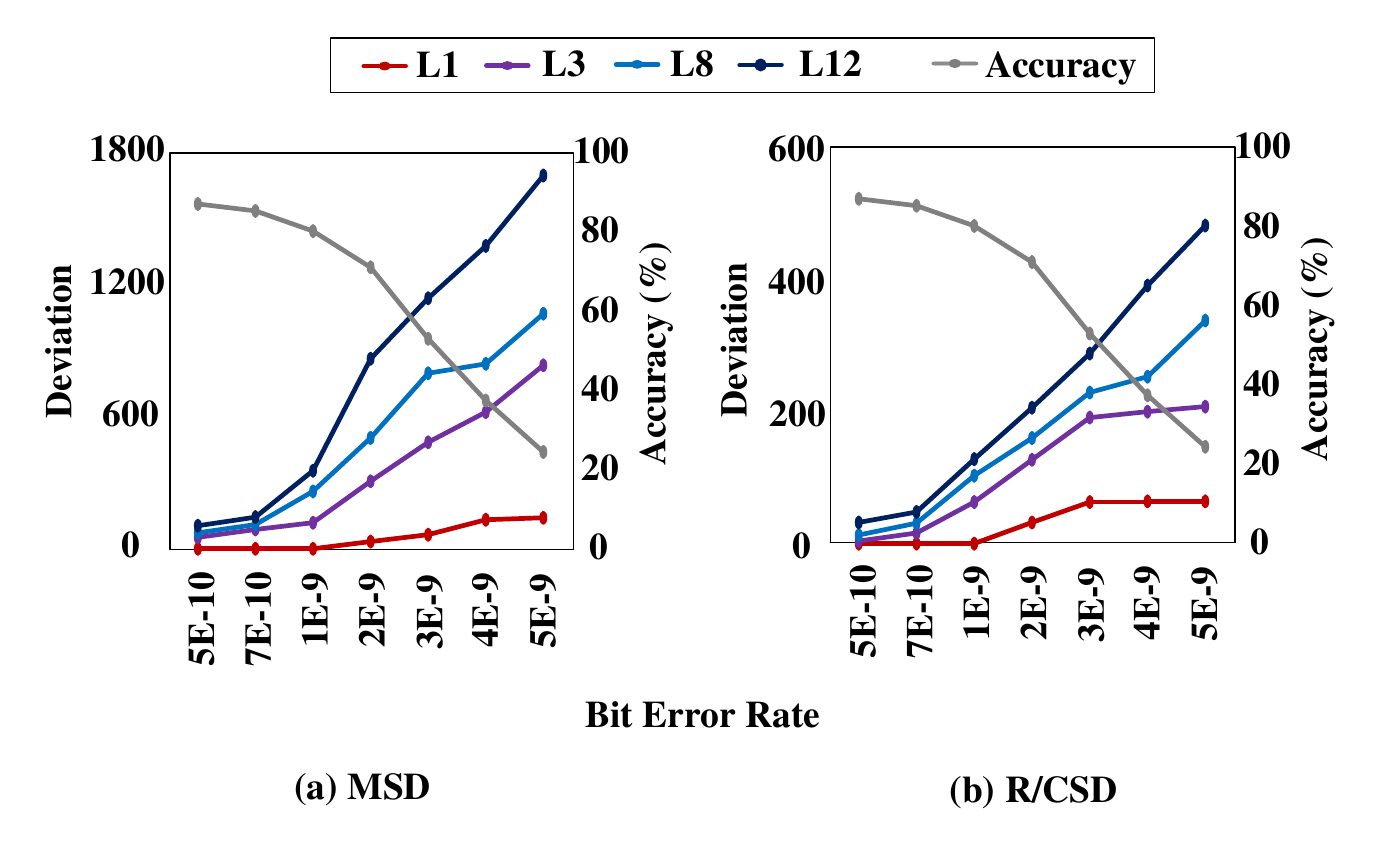}
\vspace{-0.5em}
\caption{Correlation between accuracy and mean deviation of  MSD and R/CSD.}

\label{fig:dev_ber}
\end{figure}

Fig. \ref{fig:dev_ber} illustrates the impact of MSD and R/CSD on classification accuracy across model layers under varying BER conditions, where L1, L3, L8, and L12 correspond to the 1st, 3rd, 8th, and 12th layers respectively. Experimental results demonstrate a significant correlation between the numerical distribution characteristics of MSD and R/CSD and model accuracy, despite differences in the numerical distribution of error sensitivity across different layers. Specifically, as BER increases, both MSD and R/CSD exhibit an increasing trend, forming an explicit inverse correspondence with model accuracy degradation. This finding substantiate that the values of MSD and R/CSD have a substantive impact on model accuracy.

\begin{figure}[!t]

\centering
\includegraphics[width=0.45\textwidth]{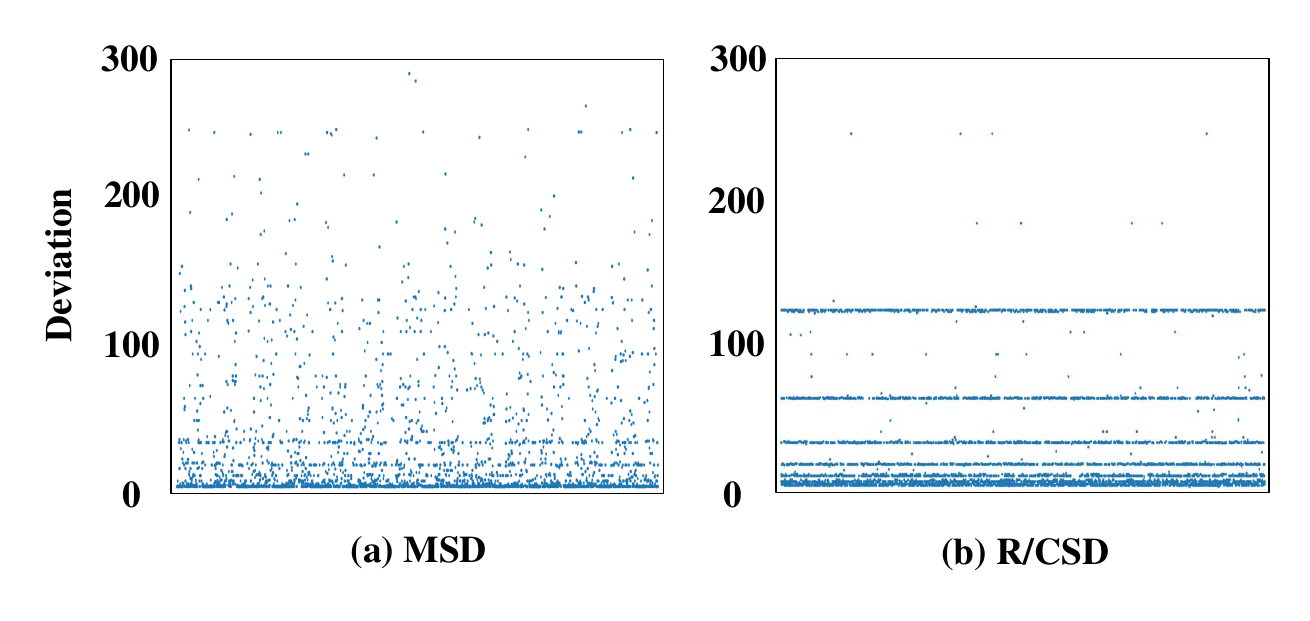}
\vspace{-0.5em}
\caption{MSD and R/CSD of the largest output matrix in each VGG19 layer.}
\label{fig:error_distribution}

\end{figure}

\begin{figure}[!t]

\centering
\includegraphics[width=0.35\textwidth]{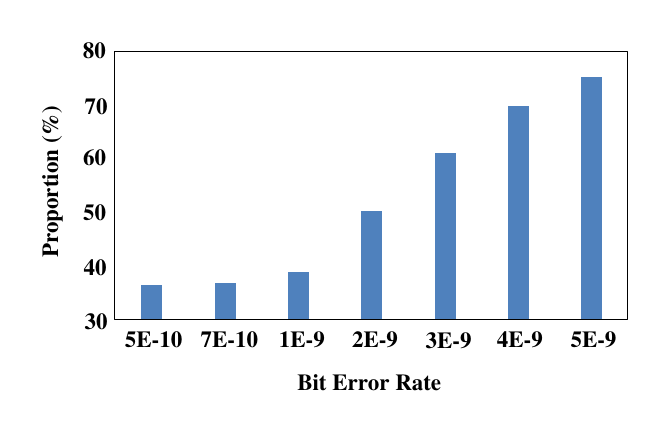}
\vspace{-0.5em}
\caption{The percentage of rows and columns that cannot be recovered by accurate ABFT due to multiple computing errors in VGG19.}
\label{fig:mult_error_proportion}

\end{figure}

To analyze the distribution characteristics of computational deviations, Fig. \ref{fig:error_distribution} presents the statistical distribution of MSD and R/CSD for the largest output matrices across VGG19 layers at BER = 7E-10. Experimental data indicate that a non-negligible proportion of near-zero MSD and R/CSD values exist, yet their actual impact on model accuracy is negligible.

Fig. \ref{fig:mult_error_proportion} quantitatively characterizes the proportion of  uncorrectable multiple computational errors occurring within rows and columns  during the implementation of previous accurate ABFT  in VGG19. Experiments reveal that as BER increases, the proportion of multi-error patterns that cannot be corrected by ABFT exhibits a significant growth trend. Furthermore, this study further uncovers the limited fault tolerance of previous accurate ABFT  mechanisms in multi-error scenarios.

Based on the aforementioned experimental observations, this research proposes an approximate ABFT optimization strategy with dynamic error thresholds. By implementing an adaptive error filtering mechanism, the proposed methodology enables dynamic suppression of negligible computational deviations. While reducing the computing overhead of error recovery, this approach compels the algorithm to correct more computational errors that have a significant impact on model output, thereby enhancing the probability of successful ABFT error recovery and effectively improving the final model accuracy.

\section{Approximate ABFT}
\label{sec:ApproxABFT}

\subsection{Overview}

\begin{figure*}[!t]

\centering
\includegraphics[width=0.9\textwidth]{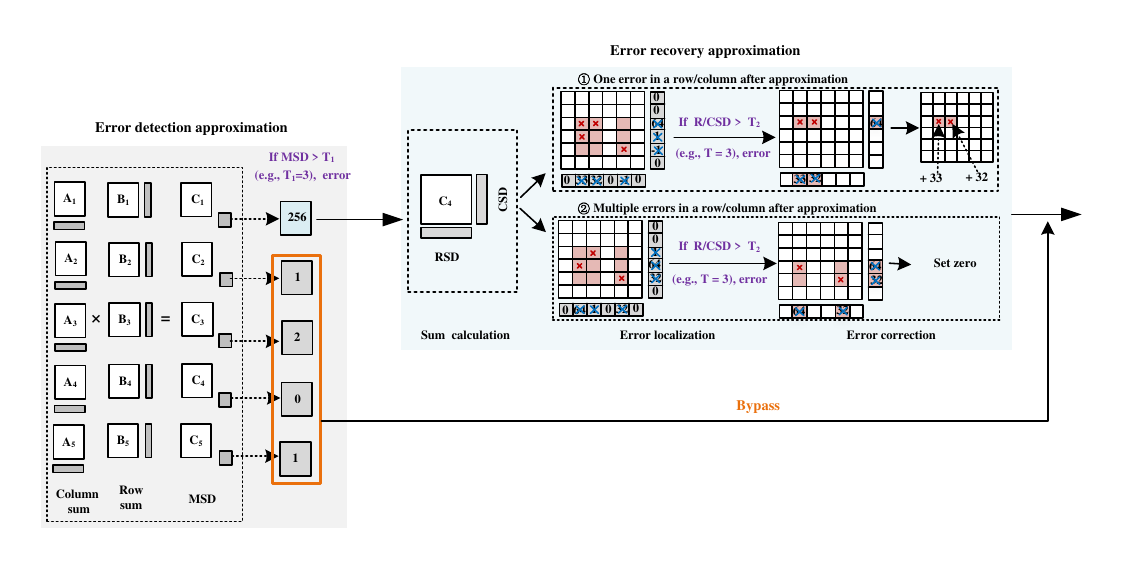}
\vspace{-0.5em}
\caption{The architecture of ApproxABFT.}
\label{fig:appro}

\end{figure*}

We propose the ApproxABFT method, which innovatively optimizes the previous accurate ABFT fault tolerance process through a dual-phase approximation mechanism. Essentially, it constructs a universal approximate verification framework that seamlessly integrates with existing matrix multiplication or convolutional operations.  

Taking the matrix multiplication-based ApproxABFT illustrated in Fig. \ref{fig:appro} as an example, during the error detection approximation phase, unlike previous accurate ABFT methods that rigidly activate error recovery procedures for all computational deviations, ApproxABFT employs an MSD threshold-guided dynamic decision framework. Specifically, the system proactively suppresses error recovery initiation when the MSD is less than or equal to the predefined threshold $T_1$. According to the MSD deviation distribution characteristics revealed in Fig. \ref{fig:error_distribution} of Section \ref{sec:motivation}, this approach can significantly reduce the number of output matrix errors that require processing, thereby optimizing the overall overhead of fault-tolerant operations. In the error recovery approximation phase, the methodology elevates the R/CSD threshold $T_2$ to filter out minor row/column deviations, thereby reducing the number of errors requiring correction. This approach converts previously uncorrectable multi-error cases into correctable single-error cases, effectively enhancing model accuracy.

\subsection{Error Detection Approximation}

Fig. \ref{fig:appro} illustrates the error detection approximation process for matrix multiplication $C= A \cdot B$, as formalized in Eq. (\ref{eq:detect}). Given matrix dimension $N$ and $\alpha$ as the N-dimensional all-ones vector $[1,1,...,1]$. The process first computes the column sum vector $A_{checksum}$ of matrix $A$ and the row sum vector $B_{checksum}$ of matrix $B$. These vectors are then used to generate the predicted checksum $C_L$ via a dot product operation. Concurrently, the actual checksum $C_R$ is obtained by summing all elements of the output matrix $C$. The MSD between the predicted value $C_L$ and the actual value $C_R$ is subsequently calculated. Based on the threshold decision mechanism defined in Eq. (\ref{eq:detect2}), the error recovery process is triggered to correct significant deviations when the MSD exceeds the predefined threshold $T_1$. Conversely, costly recovery computations are bypassed if the threshold is not exceeded.

\begin{equation} \label{eq:detect}
\begin{split}
&\mathrm A_{checksum} = \alpha A,\\
&\mathrm B_{checksum} = B \alpha ^{T},\\
&\mathrm C_{L} = A_{checksum} \cdot B_{checksum},\\
&\mathrm C_{R} = \sum_{i=1}^{N} \sum_{j=1}^{N} C_{ij},\\
&\mathrm MSD = \left | C_{L}- C_{R} \right |,\\
\end{split}
\end{equation}

\begin{equation} \label{eq:detect2}
\begin{cases}
 Invoke \; error \; recovery & \text{ if } MSD > T_{1}  \\
 Skip & \text{ if } MSD \le   T_{1}
\end{cases}
\end{equation}

The advantage of error detection approximation lies in its ability to effectively avoid error recovery operations for non-critical errors by increasing the threshold $T_{1}$. As illustrated in Fig. \ref{fig:appro}, when $T_1$ is raised from 0 (previous accurate ABFT) to 3 (ApproxABFT), the frequency of error recovery operations is reduced from 4 to 1, thereby significantly reducing the fault tolerance overhead.

\subsection{Error Recovery Approximation}

The error recovery approximation comprises three core components: sum calculation, error localization, and error correction.

The sum calculation process is detailed in Eq. (\ref{eq:sum}). First, the product of $A_{checksum}$ and matrix $B$ is computed to generate the predicted row checksum vector $C_{L\_Row}$ of the output matrix. Simultaneously, the product of matrix $A$ and $B_{checksum}$ is calculated to obtain the predicted column checksum vector $C_{L\_Col}$ of the output matrix. Subsequently, column-wise and row-wise summation operations of the output matrix  $C$ produce the actual row checksum vector $C_{R\_Row}$ and column checksum vector $C_{R\_Col}$. The R/CSD vectors are derived by calculating the deviation between the predicted and actual checksum vectors.

\begin{equation} \label{eq:sum}
\begin{split}
&\mathrm C_{L\_Row} = A_{checksum} \cdot B,\\
&\mathrm C_{L\_Col} = A \cdot B_{checksum},\\
&\mathrm C_{R\_Row} = \alpha C,\\
&\mathrm C_{R\_Col} = C \alpha ^{T},\\
&\mathrm R/CSD = \left [ \left | C_{L\_Row}- C_{R\_Row} \right |,  \; \left | C_{L\_Col}- C_{R\_Col} \right | \right ] \\
\end{split}
\end{equation}

Error localization approximation essentially validates the row/column computations of the output matrix using R/CSD, as illustrated in Fig. \ref{fig:appro}. By identifying row/column indices where R/CSD exceeds the threshold $T_2$, potential error locations in the output matrix can be precisely determined, with the coordinates of their intersection indicating the positions of potential error elements.

Error correction employs an adaptive strategy: when a row/column contains only a single error element after approximation, the corresponding R/CSD value is directly applied to the error location for accurate correction. When there are multiple error elements in the row/column, an approximate correction method is applied by resetting the error values to zero, thereby cutting off error propagation.

The advantage of error recovery approximation lies in its ability to focus on significant computational errors while ignoring minor deviations by increasing the R/CSD error metric threshold $T_2$. This strategy converts some uncorrectable error states into correctable states, thereby improving model accuracy.

\section{Block-ApproxABFT} \label{subsec:abft-block}

To further enhance fault tolerance performance in large-scale computations and multi-error concurrent scenarios, we propose a Block-based approximate ABFT (Block-ApproxABFT) mechanism. The core technology adopts a dynamic submatrix partitioning strategy, which decouples high-dimensional matrix operations into multiple independent sub-blocks and applies the ApproxABFT protection mechanism to each sub-block. This approach achieves fine-grained isolation of fault propagation paths and precise control over fault impact domains, significantly improving system fault tolerance.

\algdef{SE}[DOWHILE]{Do}{doWhile}{\algorithmicdo}[1]{\algorithmicwhile\ #1}%
\begin{algorithm}[!ht]
\footnotesize
\caption{Multi-parameter collaborative optimization algorithm}
        \hspace*{0.02in} {\bf Input:} Randomly generate $n$ sets of parameter configurations to initialize the design parameter space $S$, the number of searches is $t$. \\
\hspace*{0.02in} {\bf Output:}  Optimized parameter selection $p$.
\begin{algorithmic}[1]

\State $Bayes\_Opt()$ \Comment{Create Bayesian Optimizer}  
\State $P$ ← $\varnothing$ \Comment{Initialize the sample set}

\For {$i = 1, ..., n$}
\State $s$ ← $S_i$ \Comment{Get parameter configuration}
\State $acc$, $overhead$ ← $Get\_FT\_Val(s)$ \Comment{Get the fault tolerance accuracy and overhead under the current parameter configuration}
\State $P.append(s, acc, overhead)$ \Comment{Add new sample to P}
\EndFor 

\State $Bayes\_Opt.optimize(P)$ \Comment{Estimate posterior distribution and acquisition function}

\For {$i = 1, ..., t$}
\State $s$ ← $Bayes\_opt.select\_next\_sample()$ 
\State $acc$, $overhead$ ← $Get\_FT\_Val(s)$
\State $P.append(s, acc, overhead)$
\State $Bayes\_Opt.optimize(P)$ \Comment{Bayesian optimization to tune parameters}
\EndFor 

\State  \Return the optimized parameter selection point $p$

\end{algorithmic}  
\end{algorithm}

Given the matrix dimension heterogeneity and fault sensitivity variations across computational layers, we observe that conventional uniform parameter configuration strategies fail to achieve optimized design goals. To address this limitation, we construct a multi-parameter collaborative optimization algorithm. By dynamically adjusting critical parameters such as error detection thresholds, error recovery thresholds, and block sizes across computational layers, this algorithm establishes a trade-off mechanism between fault tolerance accuracy and redundancy overhead. To overcome the challenges of high-dimensional parameter space complexity and inefficient manual tuning, an adaptive parameter search algorithm based on Bayesian optimization is introduced. The accuracy requirement under a specified bit error rate as a design constraint to minimize fault tolerance overhead. The design space exploration problem is formalized as Eq. (\ref{eq:sep}).

\vspace{-0.5em}
\begin{equation} \label{eq:sep}
\begin{split}
&\mathrm{minimize} \quad overhead,\\
&\mathrm{s.t.} \quad acc \ge Specified \; target \; accuracy\\
\end{split}
\end{equation}
\vspace{-0.5em}

The algorithm implementation process is detailed in Algorithm 1. First, a Bayesian optimizer is constructed, and the parameter sample space is initialized, generating initial observation points within the search space (Lines 3-7). Next, the posterior probability distribution of the parameter space is inferred using the current surrogate model, and an enhanced acquisition function directs the search direction (Line 8). Finally, targeted exploration of the parameter space is conducted over t iterations (Lines 9-14), culminating in the return of the optimized parameter configuration (Line 15).

\section{Experiment Results}
\subsection{Experimental Setup}

\textbf{Datasets And Models:} 
To evaluate the effectiveness of ApproxABFT, four open-source models with varying architectural complexities and parameter scales from Table \ref{tab:table} were selected as benchmarks. Specifically, VGG19 \cite{simonyan2014very} was trained on the CIFAR10 dataset \cite{krizhevsky2009learning} containing 10 classes of 32 × 32 pixel RGB images. ResNet101 \cite{he2016deep}, DeepViT-S \cite{zhou2021deepvit}, and CaiT-XXS-24 \cite{touvron2021going} were trained on the large-scale ImageNet dataset \cite{deng2009imagenet} containing 1000 classes of 224 × 224 pixel RGB images. All neural network models employed int8 quantization to compress weight parameters.

\begin{table}[ht]

\renewcommand\arraystretch{1.3}
\caption{Major Parameters of model benchmark}

\label{tab:table}
\centering
\scriptsize

\setlength{\tabcolsep}{1.6mm}{
\begin{tabular}{|c |c |c |c |c|}
\hline
Model & Dataset & Layers & Params(M) & GFLOPs \\
\hline
VGG19 &  CIFAR10 & 16Conv+3FC & 20 & 0.8  \\\hline
ResNet101 &  ImageNet & 100Conv+1FC & 40 & 7.6 \\\hline
DeepViT-S &  ImageNet & 16Transformer & 27 & 6.2  \\\hline
CaiT-XXS-24 &  ImageNet & 24SA+2CA & 12 & 9.6  \\\hline

\end{tabular}
}

\end{table}

\textbf{Evaluation Metrics:} The fault tolerance performance of ApproxABFT is quantitatively evaluated using dual metrics of model accuracy and computing overhead. The model accuracy is evaluated using Top-1 classification accuracy. Computing overhead is represented by the additional arithmetic operations introduced by the ApproxABFT algorithm, with detailed quantification of extra multiplication, addition, and comparison operations. Assume that the input matrix dimensions are $m \times n$ and $n \times q$ respectively, let $p$ denote the error occurrence probability in matrix multiplication operations. As shown in Eq. (\ref{eq:metric}), the total computing overhead $O_{all}$ of ApproxABFT comprises error detection overhead $O_{detection}$ and error recovery overhead $O_{recovery}$. Specifically, $O_{detection}$ mainly stems from the checksum encoding overhead during the error detection process, while $O_{recovery}$ primarily involves the checksum encoding overhead required to correct erroneous matrix multiplications.

\begin{equation} \label{eq:metric}
\begin{split}
&\mathrm O_{all} = O_{detection} + O_{recovery},\\
&\mathrm O_{detection\_add} = (m-1) \cdot n + (q-1) \cdot n + (m \cdot q-1),\\
&\mathrm O_{detection\_mult} = n ,\\
&\mathrm O_{detection\_comp} = 1 ,\\
&\mathrm O_{recovery\_add} = p \cdot [(n-1) \cdot q + (n-1) \cdot m + (m-1) \cdot q],\\
&\mathrm O_{recovery\_mult} = p \cdot (n \cdot q + n \cdot m) ,\\
&\mathrm O_{recovery\_comp} =  p \cdot (m + q) \\
\end{split}
\end{equation}

\textbf{Fault Models:}  Soft errors are characterized using a bit-flip model. Specifically, based on previous reliability analysis work \cite{reagen2018ares}, the bit error rate (BER) is used as the soft error intensity metric, which is defined as the ratio of the number of bit-flip errors to the total number of bits. The BER ranges from 1E-10 to 1E-6, effectively covering the critical range of model accuracy degradation.

\textbf{Error Injection:} To evaluate the reliability of ApproxABFT, we adopt the operation-wise error injection  method proposed in \cite{xue2023exploring}, which is implemented on the PyTorch platform to inject random bit-flip perturbations into the outputs of core arithmetic operations (multiplication and addition) in neural network models. This method focuses on the computational process of the model rather than specific hardware engines, targeting the linear layers of the model architecture. We repeat the experiments for 1000 times and the confidence interval is set to be 95\%.

\textbf{Hardware Platforms:} The evaluation experiments are performed on a server equipped with two 24-core@2.5GHz Intel Xeon processors, 512GB memory, and four PH402 SKU 200 GPU cards.

\textbf{Different ABFT Setups:} Table \ref{tab:ABFT-def} compares the core configuration parameters of four fault-tolerant mechanisms: AccurateABFT, ApproxABFT, Block-AccurateABFT, and Block-ApproxABFT. The error detection phase comprises Baseline Error Detection (BED) and Approximate Error Detection (AED). The error localization phase comprises Baseline Error Localization (BEL) and Approximate Error Localization (AEL). Among them, BED and BEL employ a strict zero-threshold decision criterion. The AED threshold, AEL threshold, and block size parameter are optimized by searching within the integer interval from 0 to 100. In the error correction phase, the Baseline Error Correction (BEC) strategy ignores irrecoverable errors, the Approximate Error Correction (AEC) strategy adopts a zero-resetting operation, which effectively suppresses error propagation.

\begin{table}[ht]

\renewcommand\arraystretch{1.3}
\caption{Setups of Different ABFT Implementations}

\label{tab:ABFT-def}
\scriptsize

\setlength{\tabcolsep}{0.4mm}{
\begin{tabular}{|c|c|cc|c|}
\hline
\multirow{2}{*}{} & \multirow{2}{*}{Error Detection} & \multicolumn{2}{c|}{Error Recovery}      & \multirow{2}{*}{Block}                  \\ \cline{3-4} 
                  &                                  & \multicolumn{1}{c|}{Error Localization} & Error Correction & \\ \hline
AccurateABFT      & BED                              & \multicolumn{1}{c|}{BEL}                & BEC         & \ding{53}        \\ \hline

ApproxABFT    & AED                              & \multicolumn{1}{c|}{AEL}                & AEC      & \ding{53}         \\ \hline

Block-AccurateABFT      & BED                              & \multicolumn{1}{c|}{BEL}                & BEC         & \checkmark        \\ \hline

Block-ApproxABFT    & AED                              & \multicolumn{1}{c|}{AEL}                & AEC      & \checkmark         \\ \hline

\end{tabular}
}

\end{table}

\subsection{Overall Fault-tolerant Design Evaluation} \label{sec:overall-eval}

\begin{figure}[!t]

\centering
\includegraphics[width=0.45\textwidth]{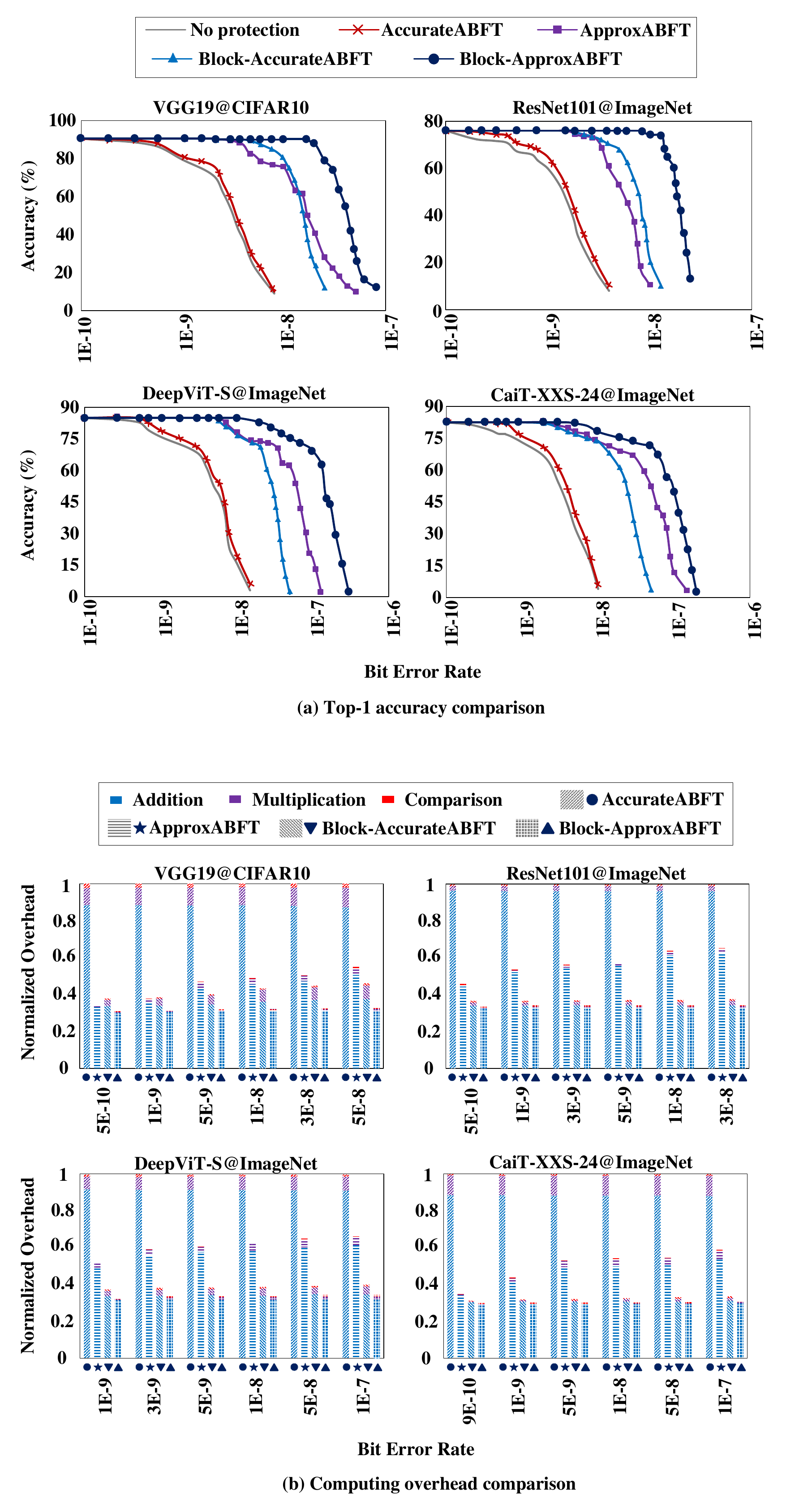}
\vspace{-0.5em}
\caption{Accuracy and computing overhead comparison between ApproxABFT, Block-ApproxABFT, AccurateABFT, and Block-AccurateABFT.}
\label{fig:AED+AEL+AEC}

\end{figure}

To systematically evaluate the performance advantages of the proposed approximate ABFT fault-tolerant design, Fig. \ref{fig:AED+AEL+AEC} compares the performance of ApproxABFT, AccurateABFT, Block-ApproxABFT, and Block-AccurateABFT listed in Table \ref{tab:ABFT-def} under different BER settings, employing two evaluation dimensions: model Top-1 accuracy and computing overhead.

1) \textit{\textbf{ApproxABFT: }}The proposed ApproxABFT was compared with the AccurateABFT fault-tolerant method. As illustrated in Fig. \ref{fig:AED+AEL+AEC}(a), the accuracy comparison reveals that AccurateABFT exhibits superior error correction capabilities when the number of computational errors is small and recoverable. In contrast, the ApproxABFT framework expands the effective BER protection range by an order of magnitude. Furthermore, normalized computing overhead comparisons in Fig. \ref{fig:AED+AEL+AEC}(b) show that ApproxABFT achieves an average 43.39\% reduction. This advantage primarily stems from the approximate error detection algorithm effectively reducing redundant error recovery operations.

2) \textit{\textbf{Block-ApproxABFT: }} The proposed Block-ApproxABFT is compared with AccurateABFT, ApproxABFT and Block-AccurateABFT fault-tolerant methods. Experimental results presented in Fig. \ref{fig:AED+AEL+AEC} demonstrate that the block partitioning mechanism significantly enhances ApproxABFT algorithms. Specifically, Block-ApproxABFT achieves superior model accuracy and lower computing overhead across various BERs, with average performance improvements of 67.83\%, 36.67\%, and 12.35\% compared to AccurateABFT, ApproxABFT, and Block-AccurateABFT, respectively. This advantage stems from two aspects: First, decomposing large-scale matrix operations into multiple sub-blocks effectively disperses random bit-flip errors, substantially reducing error density in individual sub-blocks and thereby significantly enhancing error recovery success rates. Second, the overall computing overhead is reduced by partitioning and identifying fault-free sub-blocks and applying ApproxABFT processing to the faulty sub-blocks.

\begin{figure}[!t]

\centering
\includegraphics[width=0.45\textwidth]{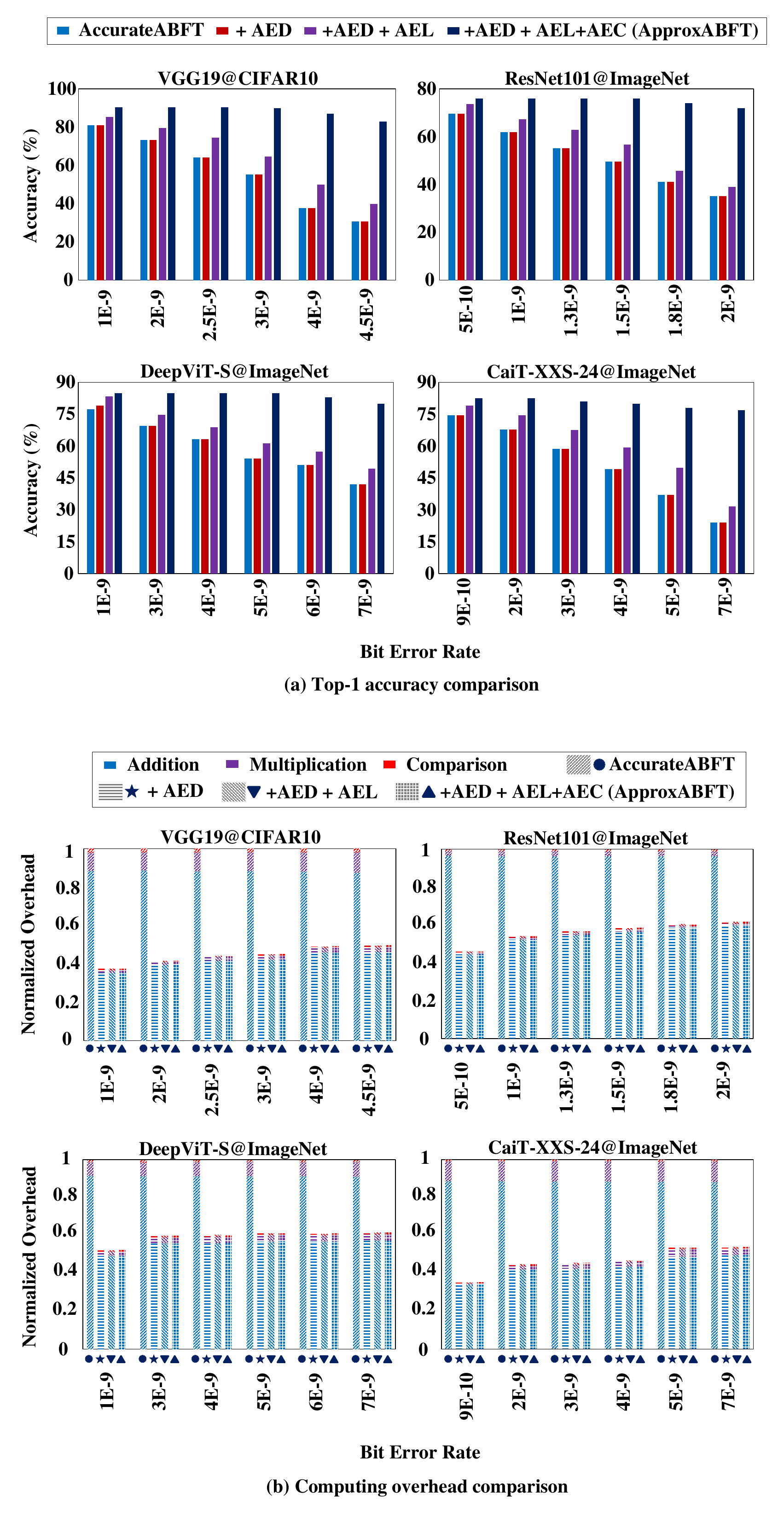}
\vspace{-0.5em}
\caption{Top-1 accuracy and computing overhead evaluation of the different approximation stages in ApproxABFT, including AED, AEL, and AEC.}

\label{fig:aedaelaec_diff}
\end{figure}

\subsection{Different Approximate Stages Evaluation}

To investigate the performance across different approximate computing stages in the proposed ApproxABFT method, Fig. \ref{fig:aedaelaec_diff} presents the comparative impacts on model accuracy and computing overhead through the progressive introduction of three key strategies: approximate error detection (AED), approximate error localization (AEL), and approximate error correction (AEC).

\begin{figure}[t]

\centering
\includegraphics[width=0.5\textwidth]{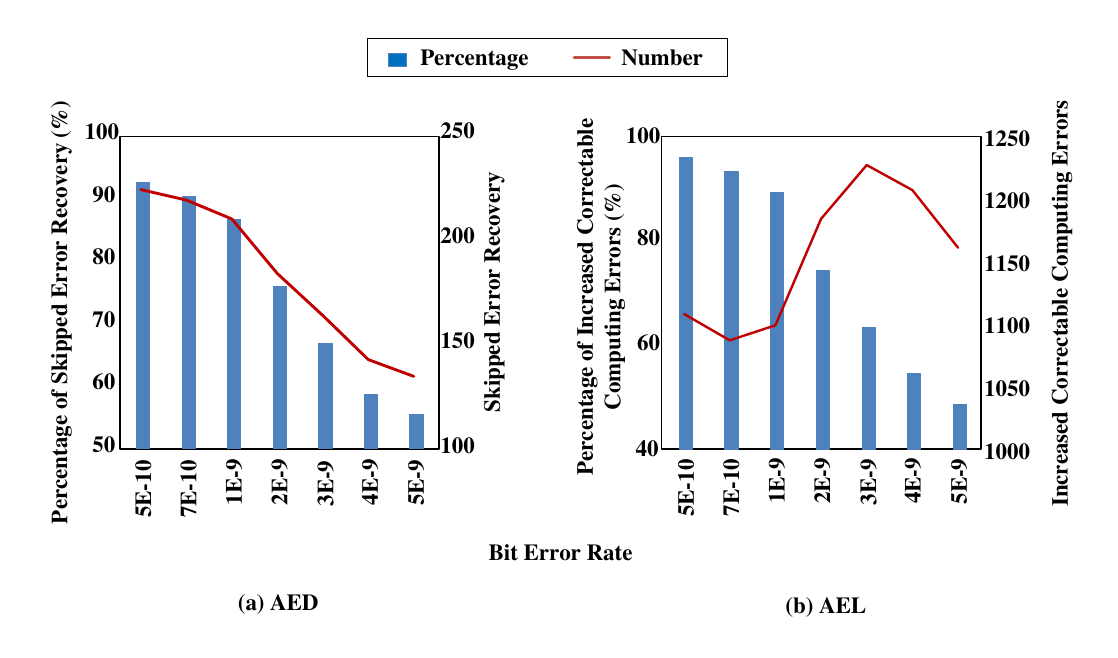}
\vspace{-1.5em}
\caption{(a) percentage of the skipped error recovery brought by AED over the total error recovery of AccurateABFT. (b) percentage of increased correctable computing errors brought by AEL over the total number of uncorrectable computing errors of AccurateABFT.}
\label{fig:probability}

\end{figure}

1) \textit{\textbf{AED:}} The experimental results in Fig. \ref{fig:aedaelaec_diff} demonstrate that the proposed AED mechanism significantly reduces computing overhead while maintaining model inference accuracy stability. Fig. \ref{fig:probability}(a) quantitatively verifies this phenomenon, which is mainly attributed to the selective avoidance of non-critical calculation errors by AED. Notably, although the proportion of avoided errors decreases with increasing BER, their absolute number remains substantial.

2) \textit{\textbf{AEL: }} The experimental results in Fig. \ref{fig:aedaelaec_diff} demonstrate that the proposed AEL mechanism improves model accuracy across varying BERs, as quantitatively evidenced by Fig. \ref{fig:probability}(b). This enhancement primarily stems from its ability to selectively  ignore minor computational errors. Consequently, ApproxABFT can more effectively rectify critical computational errors, ultimately increasing the proportion of correctable errors.

3) \textit{\textbf{AEC: }} The impact of AEC on computing overhead is negligible, yet it plays a pivotal role in enhancing model accuracy. Unlike AEL, AEC proves more effective under higher BER and numerous uncorrectable computational errors. Essentially, setting computing errors to zero has less negative impact on model accuracy because it resembles a random dropout mechanism, as validated in prior works \cite{zhang2019fault}.

\subsection{Approximation Parameter Analysis}

This subsection investigates the impact of $T_{MSD}$, $T_{R/CSD}$, and block size on model accuracy and computing overhead, and compares the proposed parameter optimization strategy with fixed parameter configuration schemes.

\begin{figure}[h]

\centering
\includegraphics[width=0.45\textwidth]{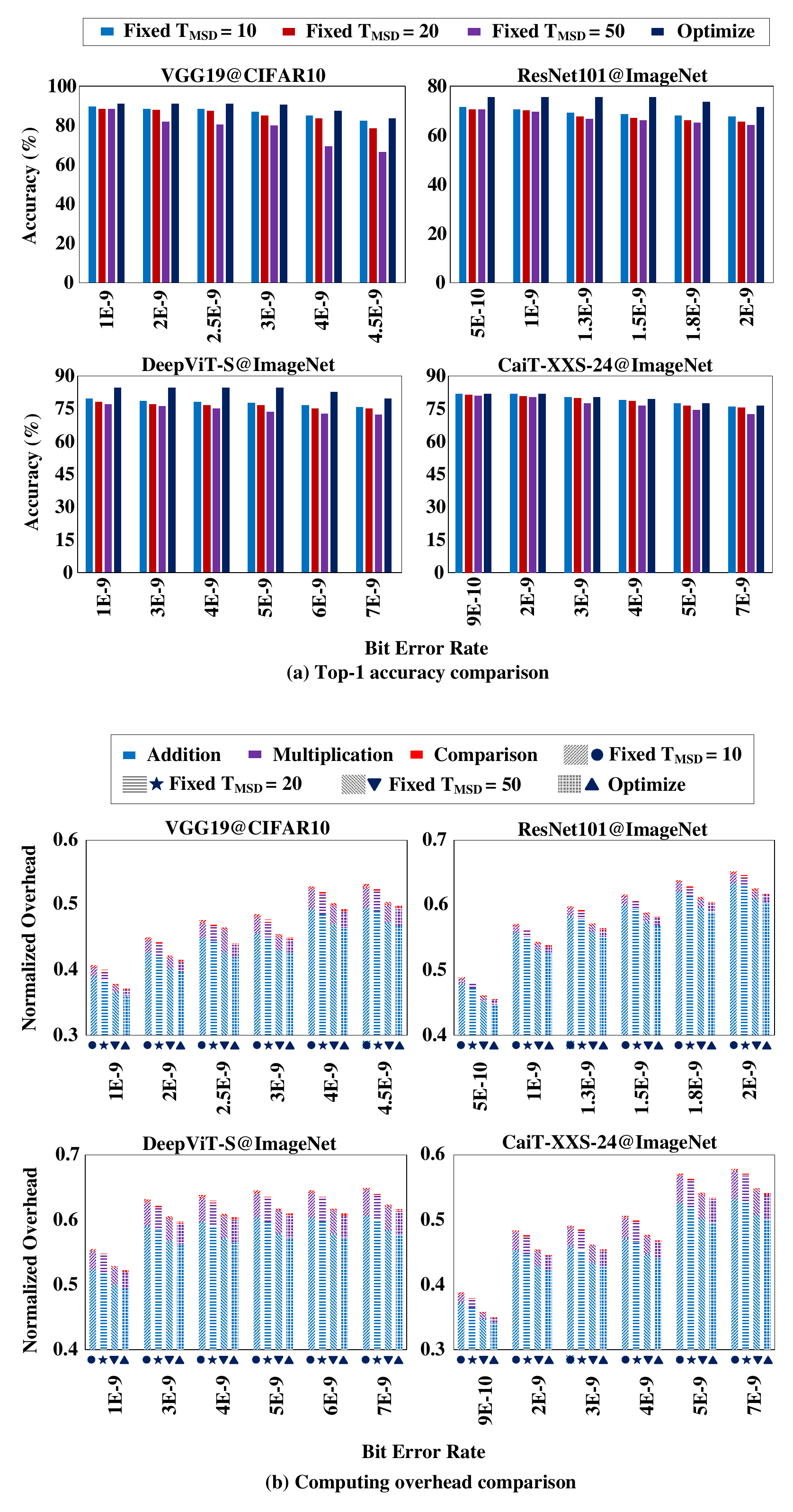}
\vspace{-0.5em}
\caption{Influence of different MSD threshold setups on ApproxABFT. Normalized computing overhead is relative to AccurateABFT overhead.}
\label{fig:diff_MSD_threshold}

\end{figure}

1) \textit{\textbf{Different MSD Thresholds: }} Fig. \ref{fig:diff_MSD_threshold} evaluates the impact of different $T_{MSD}$ threshold configurations on model accuracy and computing overhead. Experimental results demonstrate that compared with three fixed threshold settings, the optimization-based threshold configuration exhibits significant advantages in reducing computing overhead. Among the three fixed $T_{MSD}$ configurations, higher thresholds result in degraded model accuracy with diminished computing overhead. This trend primarily originates from filtering out more erroneous recovery processes, which reduces computing overhead but imposes negative impacts on model accuracy.

\begin{figure}[!t]

\centering
\includegraphics[width=0.45\textwidth]{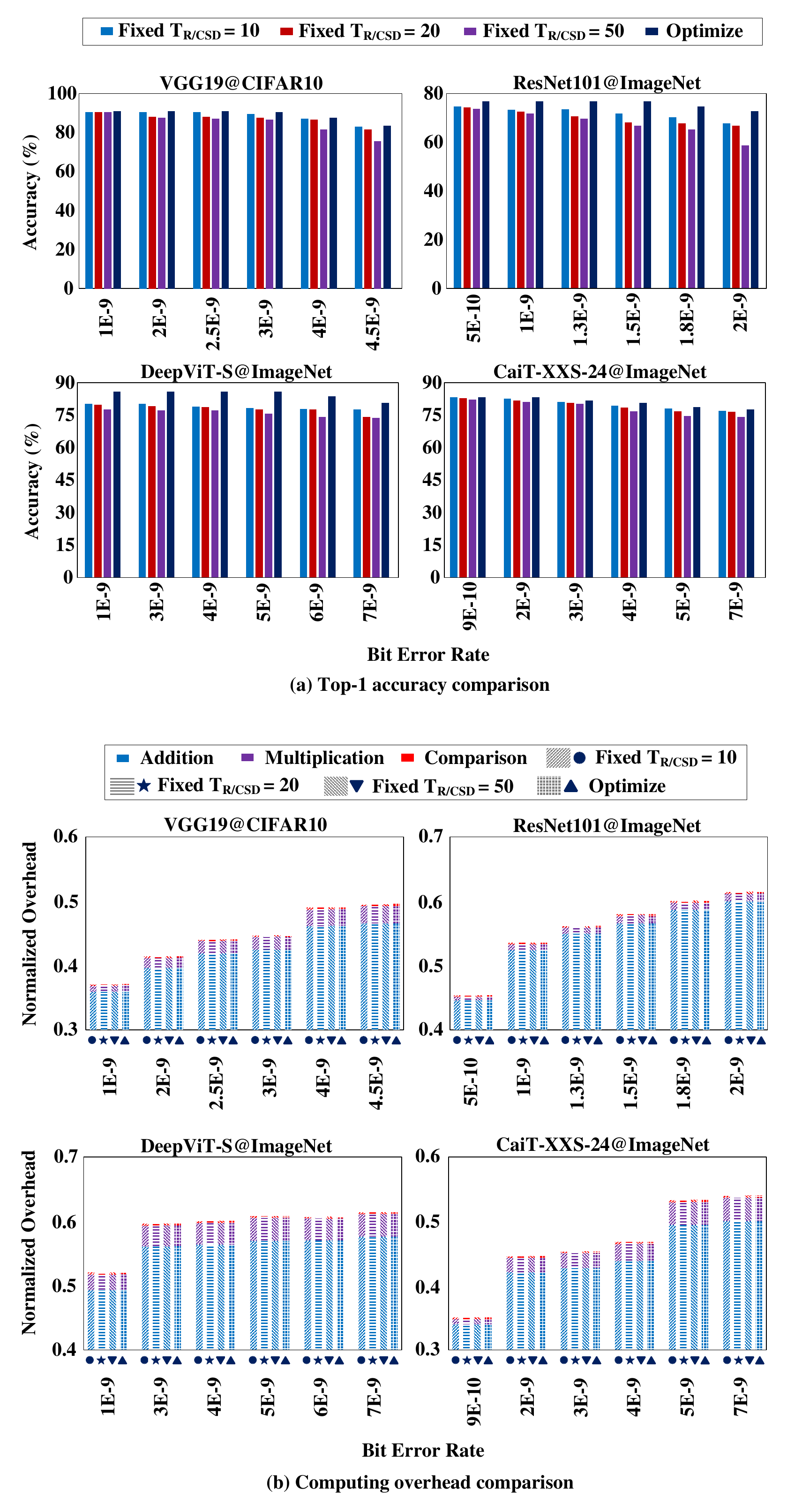}
\vspace{-0.5em}
\caption{Influence of different R/CSD threshold setups on ApproxABFT. Normalized computing overhead is relative to AccurateABFT overhead.}
\label{fig:diff_RCSD_threshold}

\end{figure}

2) \textit{\textbf{Different R/CSD Thresholds: }}  Fig. \ref{fig:diff_RCSD_threshold} compares three fixed $T_{R/CSD}$ threshold configurations with the proposed optimization-based threshold strategy. The experimental results demonstrate that while different threshold selections exhibit varying degrees of accuracy improvement, their impact on computing overhead remains negligible. Under high error rate conditions, approximation strategies can effectively enhance multi-fault error correction capabilities.

\begin{figure}[!t]

\centering
\includegraphics[width=0.45\textwidth]{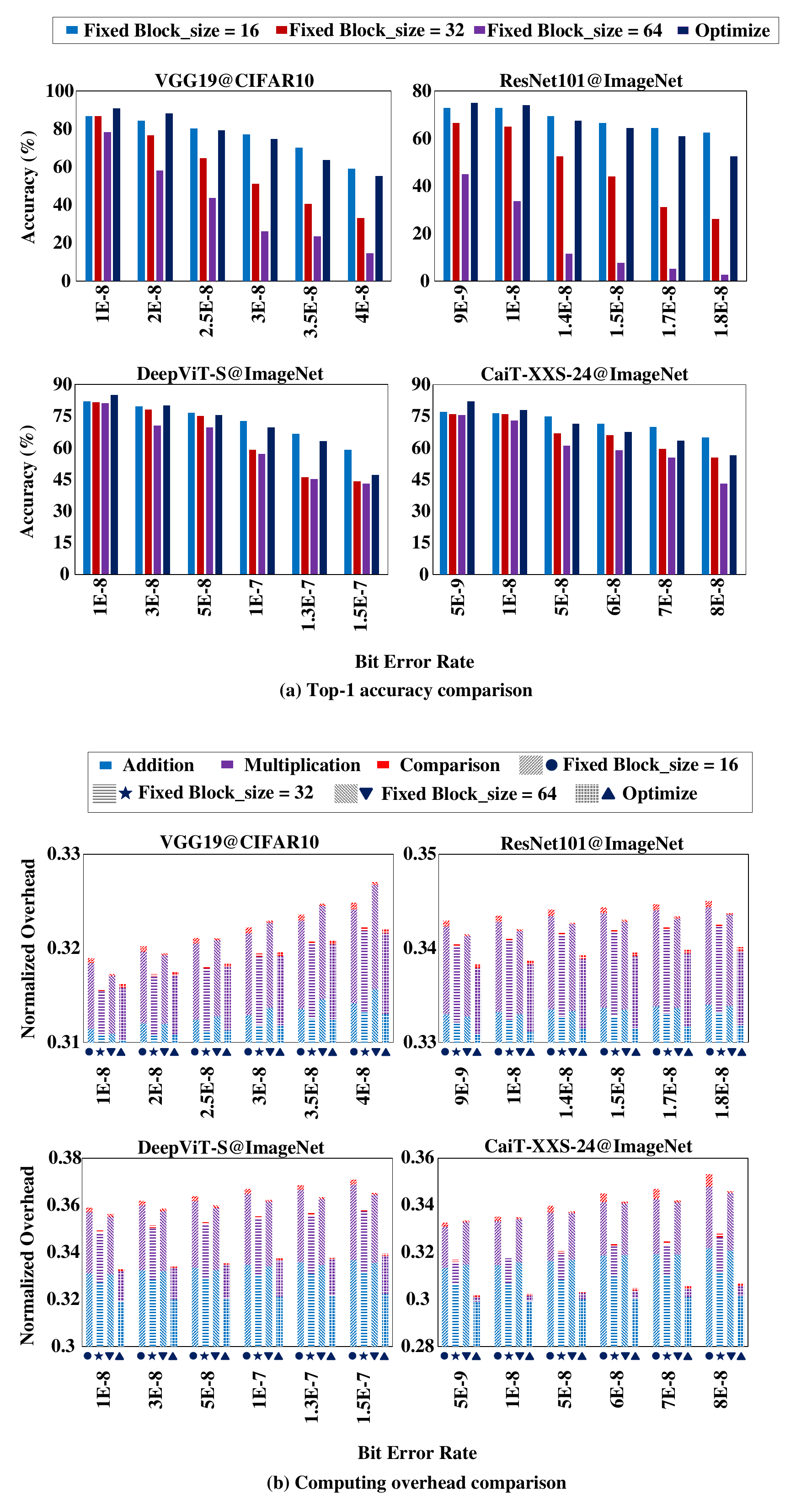}
\vspace{-0.5em}
\caption{Influence of block size setups on Block-ApproxABFT. Normalized computing overhead is relative to AccurateABFT overhead.}
\label{fig:block_size}

\end{figure}

3) \textit{\textbf{Different Block Sizes: }} Fig. \ref{fig:block_size} analyzes the impact of block size configurations on model performance. Experimental results demonstrate that smaller block sizes achieve higher model accuracy by refining the granularity of error detection and recovery. However, under specific high error rate scenarios, the accuracy exhibits slightly lower performance compared to the fixed 16-block size configuration. Regarding computing overhead, the optimized block size scheme demonstrates significant advantages across various error rate conditions. In conclusion, the optimized scheme effectively avoids increased error detection costs associated with excessively small block sizes, while simultaneously preventing the risk of error recovery failures caused by overly large block sizes.

\subsection{Comparison with Prior Work}

\begin{figure}[!t]

\centering
\includegraphics[width=0.45\textwidth]{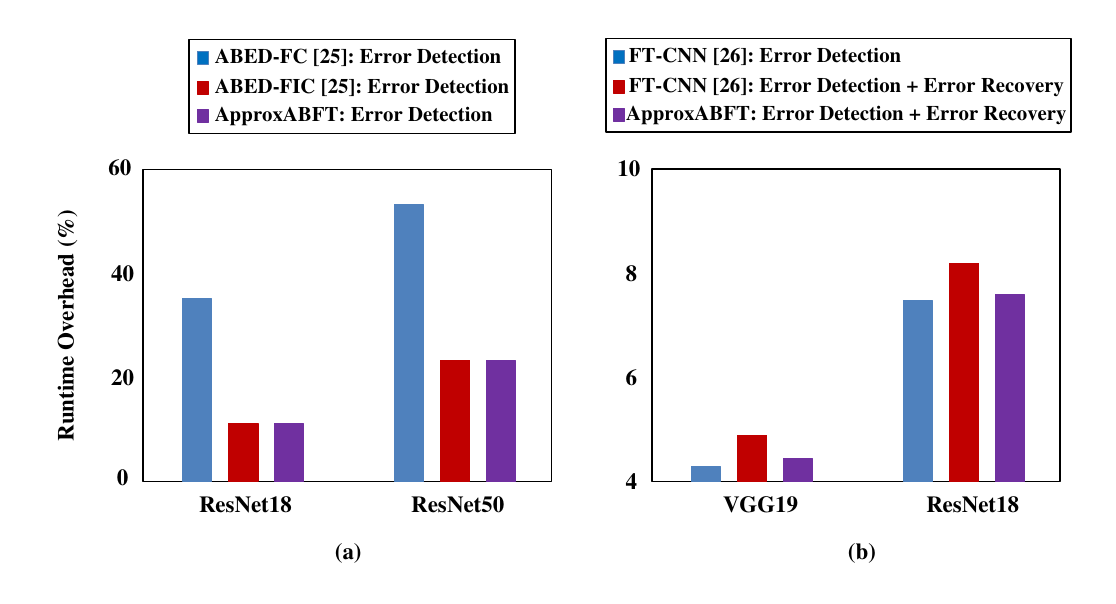}
\vspace{-1em}
\caption{(a) Comparison of the additional runtime overhead introduced by the proposed ApproxABFT and the state-of-the-art ABED \cite{hari2021making} relative to the convolution and activation operations in the model. (b) Comparison of the additional runtime overhead introduced by the proposed ApproxABFT and the state-of-the-art FT-CNN \cite{zhao2020ft} relative to the entire model.}
\label{fig:Comparison}

\end{figure}

1) \textit{\textbf{Runtime Overhead Comparison: }}  Fig. \ref{fig:Comparison}(a) compares the proposed ApproxABFT method with the FC and FIC schemes under the state-of-the-art ABED approach \cite{hari2021making} for ResNet18 and ResNet50 models on the ImageNet dataset. The evaluation specifically focuses on the additional runtime overhead introduced by the error detection mechanism relative to convolution operations and activation operations. Experimental results demonstrate that, compared with the ABED-FC scheme, the proposed method achieves a significantly reduced runtime overhead during the error detection phase. Compared to the ABED-FIC scheme, the error detection overhead of the proposed method remains comparable.

Fig. \ref{fig:Comparison}(b) compares the proposed ApproxABFT method with the state-of-the-art FT-CNN approach \cite{zhao2020ft} for VGG19 and ResNet18 models on the ImageNet dataset. The evaluation specifically focuses on the additional runtime overhead introduced by both error detection and error recovery mechanisms relative to the entire model. Experimental results demonstrate that, compared to the FT-CNN method, the ApproxABFT method effectively reduces the runtime overhead of error detection and error recovery, thereby validating the effectiveness of the proposed approach.

2) \textit{\textbf{Error correction capabilities Comparison: }}  Table \ref{tab:correct} compares the proposed ApproxABFT method with ABED \cite{hari2021making} and FT-CNN \cite{zhao2020ft} in terms of soft error correction capabilities under both low fault rate scenarios (e.g., terrestrial environments) and high fault rate scenarios (e.g., space radiation environments). Compared to ABED, the proposed ApproxABFT method significantly enhances soft error correction capabilities, effectively addressing error recovery requirements across both low and high fault rate conditions. In comparison with FT-CNN, the proposed ApproxABFT method compensates for the insufficient soft error correction capability of FT-CNN under high fault scenarios.

\begin{table}[ht]

\renewcommand\arraystretch{1.3}
\caption{Soft Error Correction Capability Comparison}

\label{tab:correct}
\centering
\scriptsize

\setlength{\tabcolsep}{0.7mm}{
\begin{tabular}{|c |c |c |c |c |}
\hline
 & ABED-FC \cite{hari2021making} & ABED-FIC \cite{hari2021making}& FT-CNN \cite{zhao2020ft} & ApproxABFT\\
\hline
Low-fault scenarios & \ding{53} & \ding{53} & \checkmark & \checkmark \\\hline
High-fault scenarios & \ding{53} & \ding{53} & \ding{53} & \checkmark \\\hline

\end{tabular}
}

\end{table}

\begin{figure}[h]

\centering
\includegraphics[width=0.45\textwidth]{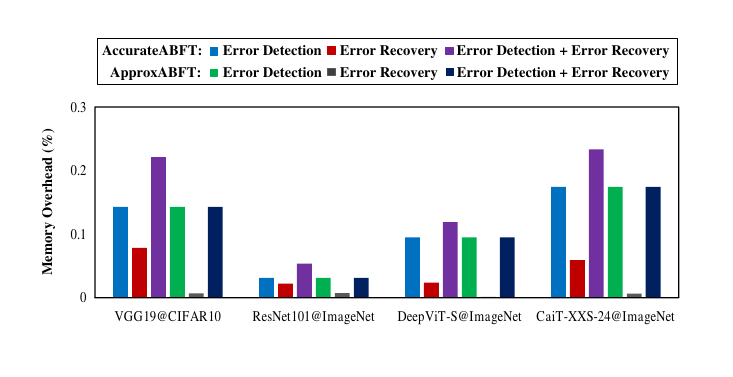}
\vspace{-1em}
\caption{Comparison of the additional memory overhead introduced by the proposed ApproxABFT and AccurateABFT relative to the entire model.}
\label{fig:memory_overhead}

\end{figure}

3) \textit{\textbf{Memory Overhead Comparison: }}  Fig. \ref{fig:memory_overhead} presents a quantitative evaluation of the additional memory overhead introduced by error detection and recovery mechanisms of the AccurateABFT and ApproxABFT methods in four typical models under single fault scenarios. Experimental results demonstrate that, compared to the AccurateABFT method, ApproxABFT exhibits no significant difference in error detection performance but achieves a notable reduction in memory overhead during error recovery. This advantage stems from the strategic relaxation of error recovery invocation by ApproxABFT, thereby minimizing redundant memory.

\section{Discussion}

The ApproxABFT method demonstrates performance advantages in both low-fault and high-fault scenarios by establishing an approximate error detection and recovery framework. Although the absolute gain is limited in conventional low-fault scenarios (e.g., terrestrial data centers), its design offers dual advantages. First, in ultra-large-scale distributed training architectures where ten-thousand-node chips form tightly coupled computational pipelines, transient errors at individual nodes risk triggering full-chain downstream failures. By intelligently filtering critical faults, ApproxABFT avoids the overreactive mechanisms of accurate ABFT, thereby reducing downstream failure propagation probabilities and minimizing unnecessary computing overhead. Second, during large-model iterative training workflows prone to frequent silent data corruption (SDC) events, ApproxABFT employs a dynamic checksum threshold adaptation strategy. By maintaining reliability while reducing system interruption rates, it significantly enhances training continuity.

\section{Conclusion}
In this paper, we propose ApproxABFT, which innovatively optimizes accurate ABFT through a two-phase approximation mechanism. The algorithm employs a multi-parameter collaborative optimization strategy to dynamically elevate error tolerance thresholds, effectively filtering out minor computational deviations. For uncorrectable computational errors, ApproxABFT utilizes lightweight zeroing operations to achieve efficient error correction. To further exploit algorithmic potential, a block partitioning strategy is integrated with ApproxABFT to adaptively adjust block granularity based on model layer dimensions. Experimental results demonstrate that, compared to previous accurate ABFT, the proposed ApproxABFT extends the effectively protected BER range by an order of magnitude, while reducing computing overhead by 43.39\% on average. 

\bibliographystyle{ieeetr}
\bibliography{ref}

\vfill
\end{document}